\documentstyle[psfig,twocolumn,aps,amstex,floats,graphicx,rotating,epsfig,amsfonts,enumerate,cite,citesort,here]{revtex}

\tighten

\begin{document}

\draft

\title{\bf Dynamic Structure Factor of
 the Two-Dimensional Shastry-Sutherland Model}  

\author{Christian Knetter and G\"otz S. Uhrig}

\address{Institut f\"ur Theoretische Physik, Universit\"at zu
  K\"oln, Z\"ulpicher Str. 77, D-50937 K\"oln, Germany\\[1mm]
  {\rm(\today)} }

\maketitle

\begin{abstract}
We calculate the 2-triplon contribution to the dynamic
structure factor of the 2-dimensional Shastry-Sutherland model,
realized in SrCu$_2$(BO$_3$)$_2$, by means of
perturbative continuous unitary transformations. 
For realistic parameters we find 
flat bound 2-triplon bands. These bands show large weight in the
structure factor depending strongly on momentum. So our findings
permit a quantitative understanding of high precision inelastic neutron
scattering experiments. 
\end{abstract}

\begin{figure}[htb]
  \begin{center}
    \includegraphics[width=5cm]{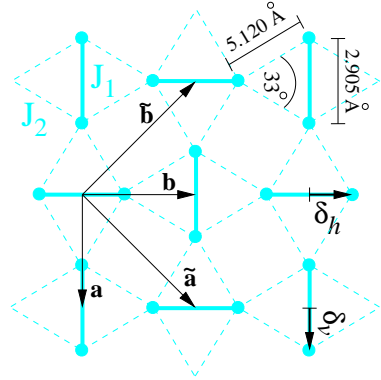}
    \caption{\label{fig:1} Shastry-Sutherland model with spin 1/2
      on the dots; dimers are solid gray lines.
     The microscopic angles and distances
      apply to SrCu$_2$(BO$_3$)$_2$ according to Ref.~\cite{smith91}. 
      The primitive vectors ${\mathbf{a}}$ and ${\mathbf{b}}$ span the dimer-lattice 
      $\Gamma_{\rm eff}$ (all dimers taken as equal) while $\tilde{\mathbf{a}}$ 
      and $\tilde{\mathbf{b}}$ span the lattice $\Gamma_{\rm AB}$ 
      distinghuishing horizontal and vertical dimers.}
  \end{center}
\end{figure}

\narrowtext
Quantum antiferromagnets which do not have a long range ordered ground
state, so-called spin liquids, continue to attract considerable
interest. While 
there are many 1-dimensional examples there are only a few 
2-dimensional systems. Of strong recent interest is
SrCu$_2$(BO$_3$)$_2$~\cite{smith91}, a realization of the
2-dimensional spin 1/2 Shastry-Sutherland model~\cite{shast81b}
\begin{equation}
  \label{H_init}
  H=J_1\sum_{<i,j>}S_iS_j+J_2\sum_{[i,j]}S_iS_j=J_1\hat U+J_2\hat V\, , 
\end{equation}
where $J_1$ and $J_2$ are the intra- and inter-dimer couplings as
in Fig.~\ref{fig:1}. Two spins coupled by
$J_1$ are referred to as dimers. We focus on $J_1,J_2>0$. 

The state $|0\rangle$ with singlets on all dimers is an exact
eigen-state of $H$ for all values of $J_1$ and
$J_2$~\cite{shast81b}. We found $|0\rangle$ to be the ground state 
(singlet-dimer phase) of
$H$ for $x:=J_2/J_1$ below $\approx 0.63$~\cite{knett00b} while
other results indicate an instability at slightly higher values
of $x$, for a review see Ref.~\cite{miyah03a}.

In many papers it has been shown that the magnetic properties of
SrCu$_2$(BO$_3$)$_2$ can
be understood well by $H$ in (\ref{H_init}) in the singlet-dimer
phase \cite{miyah03a}. Thus the borate constitutes a particularly transparent 
case of a 2-dimensional spin liquid.
In view of the extensive spectroscopic data on this system quantitative
theoretical results for spectral densities
are highly desirable. But so far only the numerical exact 
diagonalization (ED) 
for systems of  20 or 24 spins was possible \cite{miyah03a}. 
This approach  is hampered by the finite size  in two ways.
First, the energies of the excited states display strong finite size effects
since these states are spatially extended, in particular the
bound states built from two elementary excitations.
Second, the ED provides only isolated spikes instead of continuous
distributions. In this article, we remedy these drawbacks by making use
of recent conceptual progress in the method of perturbative
continuous unitary transformations (CUTs) \cite{knett03a,knett03c}. 
We provide high order results for the dynamic
structure factor which is  measured by inelastic neutron scattering
(INS).

We briefly review the derivation of perturbative
effective operators (Hamiltonian {\it and} observable) in general and
derive the appropriate INS observable for the
Shastry-Sutherland model. The corresponding
spectral density, i.e., the dynamic structure factor, is calculated via
the $T=0$ Green function. We focus on the 2-triplon part above the
 flat, featureless 1-triplon band \cite{miyah99,weiho99a,knett00e}. 

The limit of isolated dimers, i.e., $x=0$, serves as starting point of
the perturbative analysis. The basic excitation is given
by promoting one singlet to a triplet. The next higher excitation are
two triplets and so on. Upon switching on the inter-dimer coupling
($x>0$) the triplets acquire a dispersion and become dressed
particles, which we call {\it triplons}~\cite{schmi03c}. 

Up to a constant, $\hat U$ in (\ref{H_init}) counts the number 
of triplons and we define the particle-number operator $Q=\hat U+3N/4$
($N$: number of dimers). The perturbative part $\hat V$  in (\ref{H_init})
 decomposes into ladder operators $\hat V=T_{-1}+T_0+T_1$, 
where the index $i$ denotes the
number of triplons created (destroyed) by $T_i$. 

The original Hamiltonian is mapped by a perturbative CUT 
\cite{knett00a,knett00b,knett00e} to 
an effective Hamiltonian (in units of $J_1$)
\begin{equation}
  \label{H_eff}
  H_{\rm eff}(x) =
  \hat U+\sum_{k=1}^{\infty}x^k \sum_{\underline{m}}{\phantom{\Big|}}'
  C({\underline m}) T({\underline m})\, 
\end{equation}
where $\underline{m} = (m_1,m_2,\ldots ,m_k), m_i\in\{0,\pm1\}$.
This effective Hamiltonian
conserves the number of triplons: $[H_{\rm eff},Q]=0$. In
each order $k$, $H_{\rm eff}$ is a sum of virtual processes
$T({\underline m})=T_{m_1}\cdots T_{m_k}$ weighted by rational
coefficients $C$. The sum $\sum '$ is restricted by the
triplon-conservation condition $m_1+\cdots +m_k=0$. 
The effective Hamiltonian can be decomposed into irreducible  
$n$-particle operators $H_n$ \cite{knett03a}  
$H_{\rm eff}=H_0+H_1+H_2+\ldots$.

The matrix elements  of the irreducible $H_n$ for the infinite system
can be computed perturbatively on finite clusters
due to the linked cluster theorem.
For the Shastry-Sutherland model $H_0$ is conveniently set to zero; $H_1$ and
$H_2$ were determined previously \cite{knett00e,knett00b} to obtain
the 1- and 2-triplon energies. 

Applying the {\it same} transformation as for $H$ 
other observables are also mapped onto
their effective counterparts~\cite{knett03a}
\begin{mathletters}
\label{O_eff}
\begin{eqnarray}
  {\mathcal O_{\rm eff}}(x) &=& \sum_{k=0}^{\infty}x^k\sum_{i=1}^{k+1}
  \sum_{|\underline{m}|=k}\tilde{C}(\underline{m};i){\mathcal
  O}(\underline{m};i)\\
{\mathcal O}(\underline{m};i) &:=& T_{m_1}\cdots T_{m_{i-1}}{\mathcal O}
T_{m_i}\cdots T_{m_k}\ ,
\end{eqnarray}
\end{mathletters}
where ${\mathcal O}$ is the initial observable.
A useful decomposition  of ${\mathcal O}_{\rm eff}$ reads
\begin{equation}
  \label{O_decomp}
  {\mathcal O}_{\rm eff}=\sum_{n=0}^{\infty}\sum_{d\ge-n}
  {\mathcal O}_{d,n}\ ,
\end{equation}
where $d$ indicates how many particles are created
($d\ge 0$) or destroyed ($d<0$) by ${\mathcal O}_{d,n}$ whereas $n\ge 0$
is the minimum number of particles that must be 
present for ${\mathcal O}_{d,n}$ to have a non zero action. For $T=0$
measurements only the ${{\mathcal O}_{d,0}}$ operators matter.

The energy and momentum resolved $n$-particle spectral density
for the operator ${\mathcal O}$ is given by
\begin{equation}
  \label{S}
  {\mathcal S}^{(n)}(\omega,{\mathbf{K}})=-\pi^{-1}{\rm Im}{\mathcal
  G}^{(n)}(\omega,{\mathbf{K}})\ ,
\end{equation}
where ${\mathcal G}^{(n)}$ is the retarded $n$-particle Green function
\begin{equation}
  \label{G}
  {\mathcal G}^{(n)}(\omega,{\mathbf{K}})= \left\langle 0\left|{\mathcal
        O}^{\dagger}_{n,0}\frac{1}{\omega-\sum_{i=1}^nH_n+i0+} 
      {\mathcal O}^{\phantom{\dagger}}_{n,0}\right|0\right\rangle\ .
\end{equation}
Since expectation values do not change under unitary transformations
the Green function (\ref{G}) is not altered if the the effective 
operators are substituted for the initial ones.
 Using the decompositions of $H$ and of $\mathcal O$ as well as the 
conservation of triplons by $H_{\rm eff}$
 the  individual sectors of different 
triplon numbers can be analyzed separately. 

We focus on the 2-triplon sector and introduce
the 2-triplon momentum states \cite{knett00b,knett03aa}
\begin{equation}
  \label{K_states}
  |\sigma,{\mathbf{K}},{\mathbf{d}}\rangle^S =
  \frac{1}{\sqrt{N}}\sum_{\mathbf{r}}e^{i({\mathbf{K}}+\sigma {\mathbf{Q}})\cdot({\mathbf{r}}+{\mathbf{d}}/2)}
  |{\mathbf{r}},{\mathbf{r}}+{\mathbf{d}}\rangle^S\ ,
\end{equation}
where $S\in\{0,1,2\}$ is the total spin of the two triplons and
$|{\mathbf{r}},{\mathbf{r}}+{\mathbf{d}}\rangle$ is the state of one triplon on the dimer at
${\mathbf{r}}$ and the other one on the dimer at ${\mathbf{r}}+{\mathbf{d}}$. The 
 primitive vectors ${\mathbf{a}}$ and ${\mathbf{b}}$
span the dimer-lattice $\Gamma_{\rm eff}$ (Fig.~\ref{fig:1}). 
The vector ${\mathbf{K}}+\sigma{\mathbf{Q}}$ lies within the (first) Brillouin zone
(BZ) of the dual  lattice $\Gamma_{\rm eff}^*$ spanned by
the vectors ${\mathbf{a}}^*$ and ${\mathbf{b}}^*$; as usual ${\mathbf{Q}}=(\pi,\pi)$ 
in units of the inverse dimer-lattice constant. The additional quantum number
 $\sigma\in\{0,1\}$ is chosen such that ${\mathbf{K}}$ lies within the
magnetic Brillouin zone (MBZ) which is the (first) Brillouin zone of 
the dual  lattice $\Gamma_{\rm AB}^*$.
The exchange parity of the two triplons is fixed by
$|\sigma,{\mathbf{K}},{\mathbf{d}}\rangle^S=(-1)^S|\sigma,{\mathbf{K}},-{\mathbf{d}}\rangle^S$, hence
we restrict to ${\mathbf{d}}=(d_1,d_2)>0 :\Leftrightarrow \left[d_1>0\, {\rm or}\,
(d_1=0 \, {\rm and}\, d_2>0)\right]$. 

For fixed total momentum ${\mathbf{K}}$, $H_1$ (15$^{\rm th}$ order)
is a semi-infinite matrix in ${\mathbf{d}}$ and $\sigma$  while
$H_2$ (14$^{\rm th}$ order) is represented by a 84$\times$84
matrix. The matrix elements are polynomials in $x$ calculated
previously~\cite{knett00b,knett03aa}. 

We turn to analyzing the appropriate observable for the INS experiment
on the SrCu$_2$(BO$_3$)$_2$. It reads 
$
  {\mathcal F}({\mathbf{q}})=\sum_iS^z({\mathbf{x}}_i)e^{i{\mathbf{q}}\cdot{\mathbf{x}}_i}$ ,
where $S^z({\mathbf{x}}_i)$ is the $z$-component of the spin at
the position ${\mathbf{x}}_i$ (dots in Fig.~\ref{fig:1}). The
${\mathbf{x}}_i$ must not be confused with the vectors
${\mathbf{r}}\in\Gamma_{\rm eff}$ which denote the positions of the dimer centers. 

The momentum transfer ${\mathbf{q}}$ measured in  experiment is any
vector in the dual space whereas the excitations of the
Shastry-Sutherland model are  labeled best by the momenta ${\mathbf{K}}\in$ MBZ.
The usual backfolding implies ${\mathbf{K}}({\mathbf{q}})={\mathbf{q}} \mod(\Gamma_{\rm AB}^*)$
and ${\mathbf{K}}({\mathbf{q}})+\sigma({\mathbf{q}}){\mathbf{Q}}={\mathbf{q}}\mod(\Gamma_{\rm eff}^*)$ whence
$\sigma({\mathbf{q}})=0$ for ${\mathbf{q}}\in{\rm MBZ}\mod(\Gamma_{\rm eff}^*)$ and
$\sigma({\mathbf{q}})=1$ otherwise. 

We construct an operator ${\mathcal N}$ defined for ${\mathbf{K}}$ and ${\mathbf{r}}$
such, that ${\mathcal N}({\mathbf{K}};{\mathbf{r}})$ and 
${\mathcal F}({\mathbf{q}};{\mathbf{x}})$ have the same action on the ground state
$|0\rangle$. The operator ${\mathcal N}$ will then be used to
obtain the effective operator. It is a crucial feature particular to the
Shastry-Sutherland model that the triplon vacuum
$|0\rangle$ is not changed by the CUT since it is an exact eigen-state.
With a suitable convention for the
 singlet orientation the action of ${\mathcal F}$
on $|0\rangle$ is 
\begin{eqnarray}
\label{F_on_0}
  && {\mathcal F}({\mathbf{q}})|0\rangle \ = \\
  \nonumber
  &&
  i\sin({\mathbf{q}}\cdot\boldsymbol{\delta}_v)\sum_{{\mathbf{r}}^v}e^{i{\mathbf{q}\cdot\mathbf{r}}^v}
  |{\mathbf{r}}^v\rangle +
  i\sin({\mathbf{q}}\cdot\boldsymbol{\delta}_h)\sum_{{\mathbf{r}}^h}e^{i{\mathbf{q}\cdot\mathbf{r}}^h}
  |{\mathbf{r}}^h\rangle\ .
\end{eqnarray}
The sums run over all vertical dimers ${\mathbf{r}}^v$ and horizontal
dimers ${\mathbf{r}}^h$. A state $|{\mathbf{r}}\rangle$ is defined by one triplon with
$S^z=0$ on the dimer at ${\mathbf{r}}$ and singlets elsewhere. 
The vectors $\boldsymbol{\delta}_{v/h}$ are defined in Fig.~\ref{fig:1}. 

The appropriate {\it local} operator ${\mathcal N}({\mathbf{r}})$
using the distances ${\mathbf{r}}\in\Gamma_{\rm eff}$ reads
$
  {\mathcal N}({\mathbf{r}})=S_0^z({\mathbf{r}})-S_1^z({\mathbf{r}})
$,
where the 
subscripts 0 and 1 distinguish the two spins on dimer ${\mathbf{r}}$ such that
${\mathcal N}({\mathbf{r}})|0\rangle =|{\mathbf{r}}\rangle$. The
momentum space representation is given by ($\bar{\sigma}=1-\sigma$) 
\begin{mathletters}
\label{N_q} 
\begin{eqnarray}
  {\mathcal N}({\mathbf{q}}) &=& a({\mathbf{q}}){\mathcal N}(\sigma,{\mathbf{K}})+
  b({\mathbf{q}}){\mathcal N}(\bar{\sigma},{\mathbf{K}})\Big|_{\sigma({\mathbf{q}}),{\mathbf{K}}({\mathbf{q}})}{}
  \\ \label{N_qb}
  {\mathcal N}(\sigma,{\mathbf{K}}) &=& \sum_r
  e^{i({\mathbf{K}}+\sigma{\mathbf{Q}})\cdot{\mathbf{r}}}{\mathcal N}({\mathbf{r}})\\
a({\mathbf{q}})&= &i\left[\sin\left({\mathbf{q}}\cdot\boldsymbol{\delta}_v\right) +
\sin\left({\mathbf{q}}\cdot\boldsymbol{\delta}_h\right)\right]/2\\
b({\mathbf{q}})&= &i\left[\sin\left({\mathbf{q}}\cdot\boldsymbol{\delta}_v\right) -
  \sin\left({\mathbf{q}}\cdot\boldsymbol{\delta}_h\right)\right]/2\ ,
\end{eqnarray}
\end{mathletters}
which ensures ${\mathcal N}({\mathbf{q}})|0\rangle=
{\mathcal F}({\mathbf{q}})|0\rangle$ for all ${\mathbf{q}}$. 

Including the microscopic details for SrCu$_2$(BO$_3$)$_2$ (Fig.~\ref{fig:1}) and
denoting ${\mathbf{q}}$ by
${\mathbf{q}}=h\tilde{\mathbf{a}}^*+k\tilde{\mathbf{b}}^*$ fixes the scalar products
to ${\mathbf{q}}\cdot\boldsymbol{\delta}_v = 0.717(h-k)$ 
and ${\mathbf{q}}\cdot\boldsymbol{\delta}_h=0.717(h+k)$. This completes
the derivation of the excitation operator ${\mathcal N}({\mathbf{q}})$ for the INS 
 with momentum transfer ${\mathbf{q}}$.

The action of the {\it effective} local operator 
${\mathcal N}_{\rm eff}({\mathbf{r}})$ from Eq.~(\ref{O_eff}) on $|0\rangle$
 is implemented on a
computer. Although ${\mathcal N}({\mathbf{r}})$ exclusively produces 1-triplon 
states when acting on $|0\rangle$ 
${\mathcal N}_{\rm eff}({\mathbf{r}})$ leads to states containing an arbitrary
number of triplons. Focusing here on the 2-triplon channel
we have to deal with ${\mathcal N}_{2,0}$. The
calculations of the amplitudes of ${\mathcal N}_{2,0}$
can be performed on finite clusters, see Refs.~\cite{knett03c,knett03aa}.  

By substituting ${\mathcal N}_{2,0}({\mathbf{r}})$ for ${\mathcal N}({\mathbf{r}})$
in Eq.~(\ref{N_qb}) we obtain ${\mathcal N}_{2,0}({\mathbf{q}})$ which 
excites the same type of 2-triplon momentum states
$|\sigma,{\mathbf{K}}({\mathbf{q}}),{\mathbf{d}}\rangle$ that are used for the 
effective Hamiltonian. The 2-triplon amplitudes 
$
  A_{\sigma,{\mathbf{K}},{\mathbf{d}}} =
  \langle\sigma,{\mathbf{K}},{\mathbf{d}}|{\mathcal N}_{2,0}({\mathbf{q}})|0\rangle
$ with
${\mathbf{K}}\in$ MBZ
define a vector in the quantum numbers $\sigma$ and ${\mathbf{d}}$. 
Each component is calculated to 8$^{\rm th}$ order in $x$.

The 2-triplon energy and momentum resolved spectral density of
${\mathcal N}$ is obtained by evaluating the 2-particle Green
function~(\ref{G}) via
tridiagonalization~\cite{viswa94,petti85}
\begin{equation}
  \label{G_confrac}
  {\mathcal G}^{(2)}(\omega,{\mathbf{K}};x)=\frac{\sum_{\sigma,{\mathbf{d}}}|A_{\sigma,{\mathbf{K}},{\mathbf{d}}}(x)|^2} {\omega-a_0-{\displaystyle
  \frac{b_1^2}{\omega-a_1-{\displaystyle \frac{b_2^2}{\omega
  -\cdots}}}}}\ .
\end{equation}
For fixed  ${\mathbf{K}}$, the continued fraction coefficients $a_i$
and $b_i$ are obtained by repeated application of $H_1+H_2$ (matrix in
${\mathbf{d}}$ and $\sigma$) on the initial 2-particle momentum state 
$|f_0\rangle={\mathcal  N}_{2,0}({\mathbf{q}})|0\rangle$ (vector in
${\mathbf{d}}$ and $\sigma$). Prior to the evaluation
we extrapolate the matrix elements of $H_1$ and $H_2$  and the
amplitudes of $|f_0\rangle$ by optimized
perturbation theory (OPT) introduced in Ref.~\cite{knett03c}. The values
used for the OPT parameter $\alpha_{\rm OPT}$ 
are $-0.20$ for the elements of $H_1$, 
$0.80$ for the elements of $H_2$, and $-0.25$ 
for the amplitudes of $\mathcal N$ \cite{knett03aa}. 

\begin{figure}[H]
  \begin{center}
    \includegraphics[width=\columnwidth]{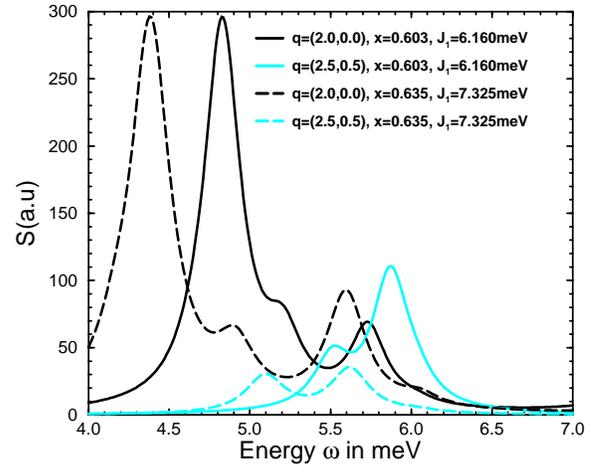}
    \caption{\label{fig:3} Two-triplon contribution to the spectral
      densities of the observable ${\mathcal N}({\mathbf{q}})$ representing 
      the INS on SrCu$_2$(BO$_3$)$_2$ for two sets of  parameters. For
      clarity, a broadening of $0.02J_1$ is used. The inaccuracy
      of the peak positions due to the extrapolation is about 2.5\%.}
  \end{center}
\end{figure}

In Fig.~\ref{fig:3}, the spectral densities of 
${\mathcal N}_{2,0}$ are plotted 
for the two momenta ${\mathbf{K}}=(0,0)$ and $(0,\pi)$,
which translate to ${\mathbf{q}}=(h,k)$ as indicated. 
Results are shown for two sets of parameters.
The set $x=0.635, J_1=7.33$ meV
was proposed  from the analysis
of the magnetic susceptibility $\chi(T)$ \cite{miyah00b}. We proposed the set
 $x=0.603, J_1=6.16$ meV previously \cite{knett00b} based on
the analysis of excitations energies at $T=0$. Clearly, the differences
between the two sets matter. Both the positions and the weights of
the curves differ from one set to the other.  Comparing to  high resolution
INS data \cite{aso02} we come to the conclusion that the parameters 
 $x=0.603, J_1=6.16$ meV fit significantly
better, both concerning the positions and
and the weights of the peaks. So we  favor this set of parameters.
It is objected that $\chi(T)$  is not well
described \cite{miyah03a}. But $\chi(T)$ is also strongly influenced by the
presence of interlayer coupling  \cite{miyah00b,knett00b} which is not
known. So a definite conclusion on the basis of $\chi(T)$ alone is 
very difficult. On the contrary, ED data for the specific heat \cite{miyah00b}
indicates that lower values of $x$ fit better to experiment than larger ones,
cf.\ the data for 16 spins.

A possible weakness in our analysis are the necessary extrapolations.
In Ref.~\cite{knett00b}, we did not use OPT but Dlog Pad\'e approximants which 
allow for power-law singularities. 
This led in a very robust way to the instability
at $x \approx 0.63$. OPT does not allow for power-law singularities.
Hence it leads to a smoother dependence on $x$.
No instability occurs below $0.7$ so that the precise position of the 
instability is still an open issue. We emphasize that in spite of the smoother
OPT extrapolation
the parameters $x=0.603, J_1=6.16$ meV still yield a better agreement with
experiment than the parameters $x=0.635, J_1=7.33$ meV. 

\begin{figure}[htb]
  \begin{center}
    \includegraphics[width=\columnwidth]{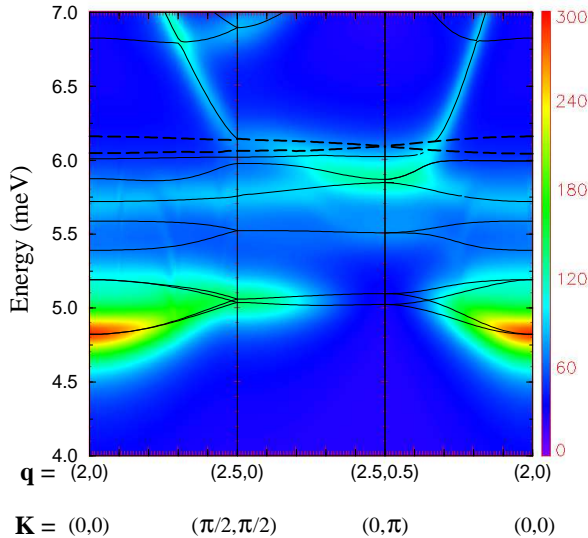}
    \caption{\label{fig:4} Color-coded spectral density of
      ${\mathcal N}$ in the energy-momentum plane. Intensities as
      indicated by the scale on the right hand side. Parameters are the
      same as in Fig.~\ref{fig:3}}
  \end{center}
\end{figure}

To understand the 2-triplon states throughout the BZ,
we calculate the spectral density for 150 different momenta. Color-coding
the intensities leads to Fig.~\ref{fig:4} where we follow the
experimentally traced path in dual space. 
The black lines are the most relevant eigen-energies of $H_1+H_2$  
extracted from the 84$\times$84
matrix representing the full matrix of $H_2$ plus a part of $H_1$
at fixed ${\mathbf{q}}$ and $x$. Enlarging this matrix from 84$\times$84
to 112$\times$112 does not lead to visible changes.
 The dashed lines mark the lower and upper bound 
 of the 2-particle continuum derived from the 1-triplon 
dispersion \cite{knett00e}. 
The  energy range depicted is chosen according to a recent high
resolution INS measurement of SrCu$_2$(BO$_3$)$_2$~\cite{aso02}. 
The 1-triplon contribution would
appear as a sharp, flat and highly intensive (red) band at about 3
meV from which no new insight is gained.

Our results compare excellently with the experimental
data. The main conclusion is that the rather flat bands of the
2-triplon states can indeed be understood. Previously, experiment
\cite{kagey00a} and theory \cite{knett00b,fukum00b,totsu01} found
evidence for significant correlated hopping of two triplons. So it
came as a surprise that high resolution INS showed very flat features
only. Previous results were limited in resolution (in momentum and in
energy \cite{kagey00a}), analysed only two points of the BZ \cite{knett00b}
or were restricted to low values of $x$  \cite{fukum00b,totsu01}.
Fig.~\ref{fig:4} shows that there are many bound states distributed over
a fairly large  energy range of about $1.5$meV. This  range
corresponds to the previous expectation of enhanced correlated hopping.
 But the smoothly 
connected eigen-energies do not display a significant dependence on momentum.
We interpret this finding as evidence for level repulsion. Due to the
negligible 1-triplon kinetic energy there is a relatively
large number of individual states involved. 
Their energetic repulsion renders each individual band very flat.

We also computed the energy and momentum integrated weights. 
At $x=0.603$, we find that about 50\%+25\% of 
the full weight, known from a straightforward sum rule,  
is covered by the 1- and 2-triplon excitations, respectively. The
remaining 25\% must be attributed to higher triplon-excitations. This
finding agrees nicely with experiment, see e.g.\ the
constant momentum scan in Fig.~2(a) of Ref.~\cite{kagey00a}.

In conclusion, we like to stress three main results. 
(i) The scenario of strongly dispersing modes in
the 2-triplon sector  cannot be
held up. We showed that these modes are also rather flat and we argue
that this stems from level repulsion.
(ii) The quantitative agreement with the
 high resolution INS~\cite{aso02} is very good. Not only the
overall shape of the structure factor but also 
prominent details like the intensive flat modes at about 5meV and the
modes just below the continuum at 5.75meV are reproduced.
 This observation supports our
choice for the parameters $x=0.603, J_1=6.16$ meV. 
(iii)
Finally, we have demonstrated that 
perturbative CUTs are capable and well-suited to
quantitatively calculate complex quantities like spectral
densities also for two-dimensional models. 

We are indebted to N.~Aso, K.~Kakurai and coworkers
for making the INS data available to us prior to publication.
Financial support by the DFG in SFB 608 and in SP 1073 is
gratefully acknowledged.

%\bibliographystyle{prsty}
%\bibliography{../../bibinput/liter10} 

\end{document}